\documentstyle[preprint,aps]{revtex}

\def\GA{\raise2.5pt\hbox{$>$}\kern-8pt\lower2.5pt\hbox{$\sim$}}
\def\LA{\raise2.5pt\hbox{$<$}\kern-8pt\lower2.5pt\hbox{$\sim$}}

\begin{document}
\draft
\baselineskip 18pt
\title{
Theory on the Itinerant Ferromagnetism 
in the $3d$-transition metal systems}
\author{Takuya {\sc Okabe}
\footnote{E-mail:okabe@ton.scphys.kyoto-u.ac.jp}}
\address{
Department of Physics, Kyoto University, Kyoto 606-01
}
\maketitle
\vskip 1cm
\begin{abstract}
\baselineskip 18pt
Keeping nickel, cobalt and iron in mind,
we investigate the origin of the itinerant ferromagnetism.
Recent experiments show that the systems should be 
in an intermediate coupling regime, where
the band width and the interaction energy are of the same order of 
magnitudes.
To treat such a situation, we generalize the Gutzwiller approximation.
In that,
we take account of the effect of 
the band degeneracy and the Hund's-rule coupling
in addition to the on-site repulsion.
In generalizing the Gutzwiller approximation to the 
bands with degeneracy,
we introduce the intuitive way to give the required expressions.
After the discussion on nickel,
the condition for
the incomplete ferromagnetism,
observed in cobalt and iron, is argued.
Phase diagrams, which show the interplay between 
the band shape peculiarity and the Hund's-rule coupling,
are given.
It is found that
for the $3d$-transition metal systems,
both of the Hund's-rule coupling and the 
special feature of the density of states are necessary 
to explain the itinerant ferromagnetism.
\end{abstract}
\baselineskip 19pt
\eject

\section{Introduction}
Recently, studies on the electron correlation are extensive.
While this is undoubtedly due to 
the high $T_{c}$ materials in addition to the heavy fermion compounds
and the related topics,
the first motivation to elucidate the nature of 
the highly correlated system was 
to clarify  the origin of the itinerant 
ferromagnetism.~\cite{rf:Kanamori,rf:Gutzwiller,rf:Hubbard}
Our purpose of this paper is to
study on the origin of the itinerant ferromagnetism
in nickel, cobalt and iron.

On the one side,
there exist a lot of works which treat
the purely mathematical models~\cite{rf:math},
since it is clear now that the effect of the electron correlation
plays an important role.
Although these works contain the interesting insights into
the mechanism for the ferromagnetism,
it may give the unrealistic impression.
On the other side,
we see the fact that the band calculation using the
local density approximation,
which includes roughly the electron correlation effects,
explains well the observed value of 
the spontaneous magnetization ${\sl etc.}$ for
the metallic ferromagnets nickel, cobalt and iron.
Still, we should note that 
we cannot identify what mechanism
plays a central role for
ferromagnetism in terms of the band calculation.
Therefore, to elucidate the physics of the
ferromagnetism,
we should start with the
model which still describes closely the real materials.
We take into account of the effect of the
band degeneracy and the Hund's-rule coupling.
As the interaction is regarded as the on-site type
in our model,
the electron correlation included in our calculation 
is thought to be included partially also in the
band calculation.
Therefore, 
our work can explain the successful result of 
the ground state property derived with use of 
the local density approximation.

Although
there exist the prominent model calculations 
since several decades ago,
most of them assumed the strong coupling interaction.
For example, Kanamori,
in his famous  work~\cite{rf:Kanamori},
argued on the magnetism of nickel,
where he considered the effect of the band degeneracy.
His conclusion was that 
we can neglect the band degeneracy and
we are allowed to treat a single band model.
This was due to his assumption $J \ll I$, where
$J$ and $I$ are respectively the exchange integral and the
interaction energy between holes with the antiparallel spins.
He also assumed $I \sim 2W$, where $W$ is the band width.
Recent experimental findings~\cite{rf:experiment} say that
these are not true but
that the Hund's-rule coupling is effective and
that the strength of the interaction is not large compared
with 
the band width.

Thus we should 
retreat the problem of the itinerant ferromagnetism
which is in
the intermediate coupling regime
of the correlated carriers in
the degenerate bands.
This is one of the motivation of our work.

Since the condition derived by Kanamori
is based on the modified Stoner
condition, $\rho(\epsilon_{F}) I_{eff} \ge 1$,
his conclusion does not necessarily show
that nickel is completely ferromagnetic in the ground state.
(It is not enough to prove the instability of the
paramagnetic state.)
Therefore, we should reinvestigate this point:
the ground state energy over the whole range of the magnetization
should be calculated 
to show the global stability of the complete ferromagnetic state.

While
Kanamori limited his argument to the case with
dilute carriers, especially for nickel,
we argue also on the incomplete ferromagnetism
which is observed in such as iron and cobalt.
Since they have more than one holes per site,
we should adopt the other approximating scheme than
the t-matrix theory by Kanamori.
Extension of  the approximation to discuss more generally 
the degenerate bands is the other purpose of
our paper.

Another aspect of our work is the reinvestigation  of 
the old ideas of Van Vleck~\cite{rf:VanVleck}
and Herring~\cite{rf:Herring}.

Some decades ago, Van Vleck~\cite{rf:VanVleck} argued that 
the effect of the Hund's-rule coupling would be
large enough to explain the itinerant ferromagnetism,
even for the
metallic nickel in which 
the hole number is so small as 0.6 per site.
According to him,
the polarity fluctuation plays an important role,
that is, more than one carriers can be on a site
at a time
to gain the kinetic energy,
and
thereby
the Hund's-rule coupling
becomes relevant.
Thus, for the Hund's-rule coupling to be effective,
the inter-particle repulsion should be reduced by the 
screening effect enough 
to allow the nearly free carrier migration.
Herring~\cite{rf:Herring}
conjectured that this actually would be the case
for nickel, and that
the band degeneracy with
the Hund's-rule coupling should be crucial
for the ferromagnetism in nickel.
(His conjecture
with respect to the effectiveness of the screening
was true in the light of the 
recent experiments.)
In other words, 
for the 3$d$ transition metal system, 
the ferromagnetism would not be realized
without the band degeneracy and the Hund's-rule coupling.
We investigate on this point.

Here in this paper,
as an approximating scheme among others, we adopt the Gutzwiller 
approximation~\cite{rf:Gutzwiller}
which is suitable for our purpose:
the procedure is physically sound 
and transparent 
in addition to
the fact that it is unrestricted by
the carrier density or the repulsion strength, 
although it is generally hard to improve on the approximation
further.
It is expected that the Gutzwiller approximation is adequate to 
estimate the ground state energy of the correlated
three dimensional systems
unless the correlation is strong enough to drive the system
toward the insulating phase.~\cite{rf:Yokoyama}
Therefore, we expect that in the range of our interest,
the approximation is legitimate.
There exist the works due to Chao and Gutzwiller~\cite{rf:Chao}
in which they generalized the Gutzwiller approximation
to the doubly degenerate bands.
However, their investigation was restricted to the 
case in the strong correlation regime,
where the events that more than two carriers are on a single
site are ignored.
Our expressions obtained below are the generalization of
their works.
Here we do not assume any restriction 
with respect to the strength of the interaction
as well as the degree of the degeneracy.

In the subsequent sections ($\S2$ and $\S3$), 
we generalize the Gutzwiller approximation to treat 
the degenerate bands.
Then, after we investigate on  the ferromagnetism in nickel ($\S4$),
we argue on the incomplete ferromagnetism ($\S5$). 
We conclude that both of the Hund's-rule coupling and the special
feature of the density of states are necessary to explain
the itinerant ferromagnetism of the 3$d$-transition metals.

\section{Gutzwiller Approximation}
As an approximating method to investigate 
the single band Hubbard model,
Gutzwiller~\cite{rf:Gutzwiller}
 made use of the variational principle.
To evaluate the expectation values 
with
the state represented by the so-called Gutzwiller 
wavefunction $|\Psi> \equiv 
\prod_i [1-(1-g)n_{i\uparrow}n_{i\downarrow}]
|\Psi_{0} >$, where
$|\Psi_{0} >$ denotes the Slater determinant
and $g (< 1)$ is the Gutzwiller variational parameter,
he introduced the further approximation, which now is known as 
the Gutzwiller approximation.
In this paper, we primarily focus on this method.
After the several illuminating studies on 
the Gutzwiller approximation,~\cite{rf:Vollhardt,rf:Ogawa}
it has become clear that
the approximation can be reproduced by a simple counting process.
Although it is simple enough that
it can be directly applicable for many models,
it still involve the great amount of task
if the model concerned is general.
Therefore, in this section,
before working on the actual variational calculation,
we give the general argument to apply the 
approximation to the general situations and 
introduce the intuitive
way to derive the expressions which are required.
Our method is simple so that
we do not have to go through the troublesome 
counting procedures.
We use neither the explicit expression
for the Gutzwiller wavefunction
nor the Gutzwiller variational parameter $g$,
but
our final expression for the ground state energy,
to be minimized, is expressed exclusively by the probabilities per site 
such as $d$, of the double occupancy.

As an example, we start with the Hubbard model.
\begin{equation}
\label{eq:H}
H=H_{kin}+H_{int}=t\sum_{<i,j>,\sigma}
a_{i\sigma}^{\dagger}a_{j\sigma}+ U\sum_{i}n_{i\sigma}n_{i-\sigma}.
\end{equation}
With the interaction $H_{int}$,
the probability that a site $i$ is doubly occupied,
 which we denote as $d_{i}$,
 is reduced from $n_{i\sigma}n_{i-\sigma}$,
the value for $U=0$.
Here $n_{i\sigma}$ designates the probability that
the $\sigma$ spin electron is on the $i$-th site.
(Or the mean number of carriers with the $\sigma$ spin.)

This reduction causes the decrease of the expectation value of
the hopping part of the energy, $<\Psi|H_{kin}|\Psi>$.
In the Gutzwiller approximation, 
this effect of the band width narrowing is taken into account
only through the factor $q_\sigma$ 
($\le 1$, independent of $i$ and $j$),
by which 
$<\Psi |a_{i\sigma}^{\dagger}a_{j\sigma}|\Psi>$ is multiplied.
This factor $q_\sigma$ should be given in terms of $d$
(independent of $i$)
and the total energy $E(d)$
 is to be minimized with respect to $d$.
\begin{equation}
\label{eq:E}
E(d)=\sum_{\sigma}q_{\sigma}(d) \tilde{\varepsilon_{\sigma}} +Ud,
\end{equation}
where
\begin{equation}
\label{eq:varepsilon}
\tilde{\varepsilon_{\sigma}} \equiv t\sum_{<i,j>}
<\Psi_0|a_{i\sigma}^{\dagger}a_{j\sigma}|\Psi_0>,
\end{equation}
is the kinetic energy of the uncorrelated state $|\Psi_0>$
for $U=0$.

We begin with rewriting $q_\sigma$ for 
$<a_{i\sigma}^{\dagger}a_{j\sigma}>$,
obtained originally by Gutzwiller~\cite{rf:Gutzwiller}, as
\begin{eqnarray}
\label{eq:q}
q_{\sigma} & = & \frac{1}{n_{\sigma}(1-n_{\sigma})}
(\sqrt{v}^{(b)}\sqrt{s_{\sigma}}^{(a)}+\sqrt{s_{-\sigma}}^{(b)}
\sqrt{d}^{(a)})_{i}
(\sqrt{s_{\sigma}}^{(b)}\sqrt{v}^{(a)}+\sqrt{d}^{(b)}
\sqrt{s_{-\sigma}}^{(a)})_{j}  \\
 & = & \frac{(\sqrt{1-n+d}\sqrt{n_{\sigma}-d}+\sqrt{n_{-\sigma}-d}
\sqrt{d})^{2}}{n_{\sigma}(1-n_{\sigma})},
\end{eqnarray}
where $n=n_{\sigma}+n_{-\sigma}$ is the number of electrons per site.
$s_{\sigma}=n_{\sigma}-d$ and $v=1-n+d$
denote the probabilities that a site is singly occupied by
the $\sigma$ spin electron and that a site is vacant respectively.
Suffices $i$ and $j$ represent the sites between which the hopping
process takes place.
The superscripts $(b)$ and $(a)$ are attached to indicate
the probability amplitudes (square root of the probability)
before and after the hopping process
respectively.
For example,
the terms $(\sqrt{s_{\downarrow}}^{(b)}
\sqrt{d}^{(a)})_i (\sqrt{s_{\uparrow}}^{(b)}
\sqrt{v}^{(a)})_j$ for $q_{\uparrow}$ and
$(\sqrt{s_{\uparrow}}^{(b)}
\sqrt{d}^{(a)})_i (\sqrt{d}^{(b)}
\sqrt{s_{\uparrow}}^{(a)})_j$ for $q_{\downarrow}$
in eq.~(\ref{eq:q})
represent the hopping process shown in Fig.1
a) and b) respectively.

The denominator takes just the same value as the numerator for
$d=n_{\sigma}n_{-\sigma}$;
$n_{\sigma}(1-n_{\sigma})$ 
might as well be written as $(\sqrt{1-n_{\sigma}}^{(b)}
\sqrt{n_{\sigma}}^{(a)})_{i}(\sqrt{n_{\sigma}}^{(b)}
\sqrt{1-n_{\sigma}}^{(a)})_{j}$.
Thus $q_{\sigma}=1$ for $U=0$ as it should be.

Here the point should be stressed:

While our derivation of eq. ~(\ref{eq:q})
( and eq. ~(\ref{eq:ql}) below)  is intuitive 
and is not justified by itself,
we can readily show that the final expressions derived after the
lengthy calculations following the usual procedure of 
the Gutzwiller approximation
can be cast into the form which can be interpreted 
as we propose here.
In effect,
the way we write down them
(which is by no means self-evident at the outset)
 is 
noticed through the course of the involved calculations.
Therefore, the approximating scheme itself does not
contain any novelty but the original idea of Gutzwiller.
However, it is a great convenience to know such
an interpretation.

Hereafter we use $d$ as the variational parameter
and the other variational 
parameter $g$ ($\eta$ in Gutzwiller's original papers)
is not considered explicitly.
It is easily seen that,
in the Gutzwiller approximation, $g$ is given by
\begin{equation}
\label{eq:g}
g^{2}=\frac{(1-n+d)d}{(n_{\sigma}-d)(n_{-\sigma}-d)},
\end{equation}
and the condition $g\le1$ is equivalent to 
$d\le n_{\sigma}n_{-\sigma}$, which is physically obvious.
It is clear from the above derivation that
$q_{\sigma}$ is independent not only of the sites $i$ and $j$
but also of the type of operators, whether creation or annihilation.
That is to say,
if $<\Psi|
a_{i\sigma}^{\dagger}a_{j\sigma}^{\dagger}|\Psi>$
(or $<\Psi|
a_{i\sigma}a_{j\sigma}|\Psi>$), for example,
 takes non-zero value for $U=0$,
in the Gutzwiller approximation,
this expectation value also
is multiplied by the
same factor $q_{\sigma}$ 
when $H_{int}$ is taken 
into account.~\cite{rf:cite}

Now, it is straightforward to generalize the above procedure.
As a result, the final expression for the total energy is given as
\begin{equation}
\label{eq:Enu}
E(\{\nu\})=\sum_{l_{1}} q(l_{1}) \tilde{\varepsilon}(l_{1})
+\sum_{p\ge 2}\sum_{\{l_{1}, \cdots , l_{p}\}}
C^{(p)}(l_{1}, \cdots , l_{p})\nu ^{(p)}(l_{1}, \cdots , l_{p}),
\end{equation}
where $\nu ^{(p)}(l_{1}, \cdots , l_{p})$
denotes the probability per site that
$p$ electrons 
occupy the $p$ states, $\{l_{1}, \cdots , l_{p}\}$ ,
with the other states being empty.
(We use the indices $l_{p}$ 
to discern the orbital as well as the spin states.)
$C^{(p)}(l_{1}, \cdots , l_{p})$
 is the intra-site repulsion energy for the 
configuration $\nu ^{(p)}(l_{1}, \cdots , l_{p})$.
The second summation sums over the different combinations
of $\{l_{1}, \cdots , l_{p}\}$.
(Note that the Pauli principle requires 
$l_{m} \neq l_{n}$ for $ m \neq n$.)
$\tilde{\varepsilon}(l)$ ($ \le 0$)
is the band energy for $C^{(p)}=0$
(in case with no correlation),
where the origin of the energy is taken such that
the completely filled band gives 
\begin{equation}
\label{eq:0}
\tilde{\varepsilon}(l)=0. \mbox{~~~~~~~~~(for the filled band)}
\end{equation}
Note that $\tilde{\varepsilon}$ is defined as a sum
over occupied states 
of the kinetic energy $\epsilon$ of each particle.
(See eqs.(2.3), (2.12) and (2.14).)
The effect of the electron correlation is included in
the factor $q(l)$, by which the band width is reduced.
As above, the factor $q(l)$ can be
written down as
\begin{equation}
\label{eq:ql}
q(l_{1})=\frac{1}{n(l_{1})(1-n(l_{1}))}
\left( \sum_{p \ge 1}\sum_{\{l_{2}, \cdots ,l_{p}\}}
\sqrt{\nu ^{(p)}(l_{1}, \cdots , l_{p})}
\sqrt{\nu ^{(p-1)}(l_{1}, \cdots , l_{p-1})} \right)^{2},
\end{equation}
\noindent
where $n(l_{1})$ is 
the average electron number in the $l_{1}$ state.
Note that $n(l_{1}) \neq \nu(l_{1})$, but
\begin{equation}
\label{eq:number}
n(l_{1})
=
\nu(l_{1})+\sum_{p(\geq2)}
\sum_{\{l_{2},\cdots,l_{p}\}}
\nu^{(p)}(l_{1},l_{2},\cdots,l_{p}).
\end{equation}

The parameters $\{ \nu^{(p)} \}$ are determined by
minimizing the energy $E(\{ \nu^{(p)} \})$, eq.~(\ref{eq:Enu}),
under the condition of the probability conservation,
\begin{equation}
\label{eq:conservation}
\sum_{p}  \sum_{\{l_{1}, \cdots, l_{p}\}}
 \nu^{(p)}(l_{1}, \cdots ,l_{p})
=1,
\end{equation}
and the number conservation, eq. ~(\ref{eq:number}), for given $n(l)$.

As for $\tilde{\varepsilon}(l)$,
it is convenient to give it as a function of the carrier density.
In principle, this is easily achieved:
\begin{equation}
\label{eq:varepsilon2}
\tilde{\varepsilon}(l)=\int^{\epsilon_{l}}\epsilon
\rho_{l}(\epsilon)d\epsilon.
\end{equation}
\begin{equation}
\label{eq:nl}
n(l)=\int^{\epsilon_{l}}\rho_{l}(\epsilon)d\epsilon.
\end{equation}
If the density of states $\rho_{l}(\epsilon)$ is given,
 $\tilde{\varepsilon}(l)$
can be regarded as an implicit function of $n(l)$
after eliminating $\epsilon_{l}$, the Fermi energy.
Here, as mentioned above (eq. ~(\ref{eq:0}))
 the origin of the energy should be given such that
\begin{equation}
\label{eq:origin}
\int^{\infty}_{-\infty}\epsilon
\rho_{l}(\epsilon)d\epsilon\equiv 0.
\end{equation}
Below we investigate on the ferromagnetism by
calculating the energy as
a function of the magnetic moment $|n_{\uparrow}-n_{\downarrow}|$,
provided that the shape of the density of states
and the total carrier density $n=n_{\uparrow}+n_{\downarrow}$
are given.

In passing, to conclude this section,
we give the simple argument to generalize the 
Gutzwiller approximation to the antiferromagnetic case.
Here, for simplicity,
 we assume a simple two-sublattice (AB) antiferromagnetic
structure.
In the Hartree-Fock approximation,
the staggered magnetization $m$ is given as 
\begin{equation}
\label{eq:m}
m( \Delta) =\frac{1}{L}\sum_{k<k_{F}}\frac{2\Delta}{\sqrt{\Delta^{2}
+\epsilon_{\bf{k}}^{2}}}.
\end{equation}
The gap $\Delta$ is given by the minimization of the energy,
\begin{equation}
\label{eq:EHF}
E_{HF}(\Delta)=\tilde{\varepsilon}(\Delta)
+\frac{U}{4} (n^{2}-m^{2}),
\end{equation}
where
\begin{equation}
\label{eq:varepsilonHF}
\tilde{\varepsilon}(\Delta)=-\frac{2}{L} \sum_{k<k_{F}}
\frac{\epsilon^{2}_{\bf{k}}}{\sqrt{\Delta^{2}
+\epsilon_{\bf{k}}^{2}}},
\end{equation}
is the kinetic energy for the antiferromagnetic state.
In the Gutzwiller approximation,
the interaction term of eq.~(\ref{eq:EHF})
should be diminished by
the reduction of $\frac{n^{2}-m^{2}}{4}=(\frac{n+m}{2})
(\frac{n-m}{2}$)$ \equiv n_{a}n_{b}$
to $d$.
As a result of this reduction,
the kinetic term should be modified
by a factor $q$.
In the same manner as above (eq.~(\ref{eq:q})),
the factor $q$
(independent of the spin component) may be written down as
\begin{equation}
\label{eq:qAF}
q=\frac{(\sqrt{1-n+d}\sqrt{n_{a}-d}+\sqrt{n_{b}-d}\sqrt{d})
(\sqrt{n_{b}-d}\sqrt{1-n+d}+\sqrt{d}\sqrt{n_{a}-d})}
{\sqrt{n_{a}(1-n_{a})n_{b}(1-n_{b})}},
\end{equation}
\noindent
where $n_{a}=\frac{n+m}{2}$ and $n_{b}=\frac{n-m}{2}$
are the number per site
of 
the electrons with the spin component parallel
and antiparallel to the staggered magnetization, respectively.
($q=1$ for $d=n_{a}n_{b}$ as expected.)
In consequence,
the variational parameters are $\Delta$ and $d$,
and are determined by minimizing the energy,
\begin{equation}
\label{eq:EAF}
E(\Delta,d)=q\tilde{\varepsilon}(\Delta)+Ud
\end{equation}
Note that
the above expressions are the same as given by
Kotliar and Ruckenstein~\cite{rf:Kotliar},
and Metzner~\cite{rf:Metzner},
who solved the long-standing problem of generalizing the 
Gutzwiller approximation in a satisfactory manner
to antiferromagnetism.

To conclude this section, we repeatedly stress that
our intuitive derivation of the factor $q$ is 
based on the laborious but simple counting procedure.
Our interpretation was given for 
convenience' sake as a hindsight.

\section{Formulae}
Neglecting the orbital dependence of the interaction energy,
we parameterize the on-site interaction only by 
the two parameters $I$ and $I-J$,
which represent the intra-site repulsion between the two carriers
with the antiparallel and the parallel spin components, respectively.
The Hund's-rule coupling is taken
into account as the exchange integral $J$.
For the bands with the $D$-fold degeneracy,
we assume that the kinetic energy of the uncorrelated band
can be written as $D\tilde{\varepsilon}(\frac{n_{\sigma}}{D})$,
which means that all of the $D$ bands have the same structures.
As for the filling,
we assume no orbital ordering~\cite{rf:orbitalorder}
and regard each of $D$ orbitals as equivalent.
In this case, the expression for the total energy can be written down 
in terms of the parameter $\nu_{p,q}$
which denotes the probability 
for each of the configurations which has the $p$ up-spin 
and $q$ down-spin electrons on a site.
(Compare with $P(p,q)$, below eq.~(\ref{eq:P0}).)

\begin{equation}
\label{eq:Enu2}
E(\{\nu\})=\sum_{\sigma=\uparrow,\downarrow} 
q_{\sigma} D\tilde{\varepsilon}(\frac{n_{\sigma}}{D})
+\sum_{p,q (p+q \ge 2)}^{p+q=2D}
{_{D}C_{p}}\cdot {_{D}C_{q}}\nu_{p,q} \left \{pqI
+({_{p}C_{2}}+{_{q}C_{2}})(I-J) \right \},
\end{equation}
\noindent
where
\begin{equation}
\label{eq:qup}
q_{\uparrow}=
\frac{1}{\frac{n_{\uparrow}}{D}(1-\frac{n_{\uparrow}}{D})}
\left(
\sum_{p=1 ,q=0}^{p+q=2D}    
{_{D-1} C_{p-1}} \cdot {_{D} C_{q}} \sqrt{\nu_{p,q}}\sqrt{\nu_{p-1,q}}
\right) ^{2},
\end{equation}
\noindent
and
\begin{equation}
\label{eq:qdown}
q_{\downarrow}=
\frac{1}{\frac{n_{\downarrow}}{D}(1-\frac{n_{\downarrow}}{D})}
\left(
\sum_{p=0 ,q=1}^{p+q=2D}
{_{D} C_{p}} \cdot {_{D-1} C_{q-1}} \sqrt{\nu_{p,q}}\sqrt{\nu_{p,q-1}}
\right) ^{2}.
\end{equation}
\noindent
(We used the notation $ {_{N}C_{M}} \equiv \frac{N!}{M!(N-M)!}$,
which counts the combination of the orbitals with the same energy.)

The conservation laws 
(eqs.~(\ref{eq:number}) and ~(\ref{eq:conservation})) 
require
\begin{equation}
\label{eq:nu10}
\nu_{1,0}=
\frac{n_{\uparrow}}{D}-\sum_{p,q (p+q \ge 2)}^{p+q=2D}
{_{D-1} C_{p-1}} \cdot {_{D} C_{q}} \nu_{p,q},
\end{equation}
\begin{equation}
\label{eq:nu01}
\nu_{0,1}=
\frac{n_{\downarrow}}{D}-\sum_{p,q (p+q \ge 2)}^{p+q=2D}
{_{D} C_{p}} \cdot {_{D-1} C_{q-1}} \nu_{p,q},
\end{equation}
and
\begin{equation}
\label{eq:nu00}
\nu_{0,0}=1-n+\sum_{p,q (p+q \ge 2)}^{p+q=2D}
(p+q-1) {_{D}C_{p}}\cdot {_{D}C_{q}}\nu_{p,q},
\end{equation}
where $n=n_{\uparrow}+n_{\downarrow}$.
Therefore,
eliminating $\nu_{1,0}, \nu_{0,1}$ and $\nu_{0,0}$ 
by eq. ~(\ref{eq:nu10}), ~(\ref{eq:nu01}) and ~(\ref{eq:nu00}),
we should minimize eqs. ~(\ref{eq:Enu2}) with respect to 
${ \nu_{p,q} }$ with $p+q \ge 2$.
After elimination of $\nu_{p,q}$,
the energy $E$ can be regarded as a function of 
$n_{\uparrow}$(or $n_{\downarrow})$.
We then seek for the minimum of $E(n_{\uparrow})$ to study
magnetism.

We remark here that the probability $P(p,q)$ that
a site is occupied by the $p$ up-spin and $q$ down-spin 
electrons is given by
\begin{equation}
\label{eq:P0}
P(p,q)=
{_{D}C_{p}}\cdot {_{D}C_{q}}\nu_{p,q}.
\end{equation}
In the case where $C(p,q)=0$ for all $p$ and $q$,
we can give $P(p,q)$ explicitly by
\begin{equation}
\label{eq:P}
P(p,q)={_{D}C_{p}}\cdot {_{D}C_{q}}
\left(\frac{n_{\uparrow}}{D}\right)^{p}
\left(\frac{n_{\downarrow}}{D}\right)^{q},
\end{equation}
which is the expression that Van Vleck~\cite{rf:VanVleck}
used to estimate the degree of 
the polarity fluctuation.
Then, it is easily seen that
\begin{equation}
\label{eq:Psum}
\sum_{p,q \ge 1}P(p,q)=1,
\end{equation}
as expected. (eq. ~(\ref{eq:conservation}.))

\section{Nickel}
Although we studied the itinerant ferromagnetism in nickel
previously~\cite{rf:nickel},
our consideration there was
under the restriction that 
only up to two carriers can sit on the same site
at most
as in the work of Chao~\cite{rf:Chao}
(In our notation above, only 
$\nu_{2,0}$, $\nu_{1,1}$ and $\nu_{0,2}$
were taken into account as variational parameters.)
While this is not a bad assumption in view of the smallness of 
the carrier density, 0.6 per site,
we can improve upon in this respect in terms of the results
of the previous section.
The other reason for the
reinvestigation for nickel here 
is that there we overestimated the
value of the Hund's-rule coupling $J$.
Here in this paper, as parameters for nickel, 
we use~\cite{rf:experiment}
 $n\sim0.6$eV, $W\sim4$eV, $I\sim$2.4eV and $J\sim$1.2eV,
 i.e. , $i \equiv I/W \sim 0.6$ and $j \equiv J/W \sim 0.3$.
Anyway, the result remains the same:
in the ferromagnetic state,
about $10\%$ of the probability that
a site has two carriers (holes)
($\sum_{p,q,p+q=2}P(p,q)=P(2,0)+P(1,1)+P(0,2)\sim 0.1$
for $I=J=0$),
is not so modified by including the effect of $I$ and $J$.
That is to say,
the on-site interaction energies to be
adopted in our calculation has already been
screened by conduction electrons in such a manner that
$I$ is reduced to be
comparable with the band width $W$ but keeping the
effectiveness of $J$ nearly intact.~\cite{rf:experiment}

To discuss the ferromagnetism of nickel and palladium,
Kanamori~\cite{rf:Kanamori} took account of the fact that
the holes in the $d$ band occupy states of the Brillouin zone 
close to the point $X$, where the symmetry of the band-edge state is 
$X_{5}$.
The six $X_{5}$ states (two degenerate states at each of the three 
equivalent $X$ points) are made up of the 
three atomic functions of the $t_{2g}$ class (
$xy$, $yz$ and $zx$ symmetries).
Consequently, 
we regard the system as consisting of the triply degenerate bands,
i.e., $D=3$ as Kanamori.
Then we have 13 variational parameters, $\nu_{p,q}$ with 
$2 \le p+q \le 2D$.
As for the kinetic energy,
it is known that 
the density of states 
shows a significant peak near the top of the band
at the Fermi energy
for a hypothetical nonmagnetic nickel and also for palladium.
Kanamori argued that
the quantitative difference of the
density of states at the Fermi energy
is one of the reasons
by which palladium remains paramagnetic
while the ferromagnetism is realized in nickel.
Since 
it is clear that such a specific feature of 
the density of states
is important,
we take this into account by
the
schematic density of states 
$\rho$
as a function of the carrier number per state,
$n$, as
\begin{equation}
\label{eq:nidos}
\rho(n)= \left\{
\begin{array}{ll}
\displaystyle{C_1}&
\qquad 1 \ge n \ge 2n_0 \\
\noalign{\vskip0.2cm}
\displaystyle{\frac{C_1}{r+(1-r)(\frac{n-n_0}{n_0})^2} }&
\qquad 2n_0 \ge n \ge 0
\end{array} \right.  . 
\end{equation}
The Fermi energy 
$\epsilon$ as a function of $n$ can then be given as
\begin{equation}
\label{eq:epsilonn}
\epsilon(n) = \int^{n}_{0} \frac{1}{\rho(n)} dn +C_2 .
\end{equation}
The normalization constant $C_1$
is determined by the requirement that
\begin{equation}
\label{eq:normalization1}
\epsilon(1)-\epsilon(0) \equiv W,
\end{equation}
by the definition of the band width $W$.
And the origin of the energy $C_2$ is given such that
\begin{equation}
\label{eq:normalization2}
\int^{\infty}_{-\infty} \epsilon(n)dn=0,
\end{equation}
according to eq. ~(\ref{eq:origin}).
The kinetic energy $\tilde{\varepsilon}$ is now 
given as a function of $n$ by
\begin{equation}
\label{eq:varepsilonformula}
\tilde{\varepsilon}(n)=\int^{n}_{0}
\epsilon(n) dn.
\end{equation}

In Fig.2  and Fig.3,
we show $\rho$ as a function of $n$ and $\epsilon$,
where we set $n_0=0.1$, $W=1$ and $r^{-1}=2$.
Below we fix $n_0=0.1$ so that the Fermi energy of the
hypothetical paramagnetic nickel is just at the peak,
and use the relative height of the peak, $r^{-1}$, 
as a parameter which characterizes the density of states
of nickel.

In Fig.4
 we show the phase diagram for $j \equiv J/W =0.3$
obtained with use of the Gutzwiller approximation,
together with the results of the Hartree-Fock approximation.
Compared with the latter,
we see that the paramagnetic phase is stabilized remarkably.
This reflects the fact that 
the one-body approximation should not be taken seriously 
with regard to the appearance of the ferromagnetism.
We also see that the phase boundaries
 for the Gutzwiller approximation
are almost independent of $i\equiv I/W$.
This is because of the fact
that 
the strength of the effective interaction 
(which cannot exceed the band width) does not depend 
strongly on the bare one 
in the strong coupling regime.
In a way, this can be taken as representing the fact that
the peak at the density of states, the factor $r^{-1}$
rather than the Hund's-rule coupling $J$,
plays decisive role to realize ferromagnetism in nickel.
However, we should note that it is due to the
Hund's-rule coupling effect which makes
the reasonable range of $r^{-1} \sim 2$
to be the complete ferromagnetic region.
In Figs.5 and 6,
 we show the phase diagrams for $j=0.15$ and $j=0$,
 respectively.
Without the Hund's-rule coupling,
or $J=0$,
the argument essentially renders back to the case for
the non-degenerate band,
for which it is known that
ferromagnetism is hard to be realized in
a realistic range of parameters. 
Note that the condition derived by Kanamori 
corresponds to the dotted curve, which 
is the boundary between the paramagnetic
 and the incomplete ferromagnetic state,
since
Kanamori's argument is based on the modified Stoner
condition, $\rho(\epsilon_{F}) I_{eff} \ge 1$.
In this respect,
his conclusion for
nickel
should be re-investigated
because actual nickel is completely ferromagnetic.
We see that, to obtain the complete ferromagnetic state
without the Hund's-rule coupling,
$r^{-1}$ should be larger than 5 even for $i \sim 1$.
It is,  therefore, legitimate to say that
the Hund's-rule coupling plays an decisive role
for the complete ferromagnetism.

We should also note on the continuous transition from
paramagnetism to complete ferromagnetism.
Generally speaking, the boundary between the complete
and incomplete ferromagnetism is independent of 
that between 
the incomplete ferromagnetism and the paramagnetism.~\cite{rf:cite2}
Therefore the two boundaries can cross at some strength of the
interaction.
Dashed curve in Fig.4
represents the first order transition
line around which the both of the complete ferromagnetic and
the paramagnetic state can be locally stable.
Such a boundary arises in the region of the
strong interaction, that is to say,
likely to appear without the Hund's-rule coupling
nor the peculiarity of the density of states.

In Fig.7,
 we show the magnetic moment $|n_{\uparrow}-n_{\downarrow}|$
as a function of $n \equiv n_{\uparrow}+n_{\downarrow}$
for $i=0.6$, $i-j=0.3$, $n_{0}=0.1$ and $r^{-1}=2.4$.
We see that the complete ferromagnetism
appears for 
the values of $n$ for which
the Fermi surface lies in the peak portion,
while 
above or below some definite values of $n$
( $n\GA1.01$ and $n\LA0.27$) 
the state becomes unstable to 
result in the precipitous decrease of the magnetic moment.

As for the factor $q$,
since the electron correlation is not strong,
the band width does not change drastically.
For example, for $i=0.6$, $i-j=0.3$, $n_{0}=0.1$ and $r^{-1}=2.4$,
the factor $q$ is 0.98 for the complete ferromagnetic state
while $q=0.94$ for the hypothetical paramagnetic state.
Accordingly, the probabilities, eq.~(\ref{eq:P0}), 
obtained variationally 
do not
modified appreciably from those for the  uncorrelated
 values~\cite{rf:nickel}, 
eq.~(\ref{eq:P}).
If we approximate the exchange splitting $\Delta$ by 
$|\frac{{\rm d}E(n_{\uparrow})}{{\rm d}
n_{\uparrow}}|_{n_{\uparrow}=0}$,
which is valid if the carrier density $n$ is
small,
we can compare the result with the Hartree-Fock and the Gutzwiller
approximation. The result is 
$\Delta_{\rm HF}=3.65\Delta_{\rm Gutzwiller}$
for the same parameters as above.

\section{Incomplete Ferromagnetism}
In this section, we investigate the incomplete ferromagnetism,
which is observed in cobalt and iron.
To the author's knowledge,
the quantitative estimation of the
parameters such as the Hund's-rule coupling
are not known experimentally.
Therefore, noting the fact that
the strength of the Hund's-rule coupling would not drastically 
change 
from nickel to iron,
we proceed to see what should be the
condition which makes the complete ferromagnetic state
unstable, while keeping the paramagnetic state also unstable.
Here we treat the hypothetical metal of the 
bands with the five-fold degeneracy, $D=5$,
with the carrier density $n=2.0$,
keeping in mind the metallic cobalt and iron.
As variational parameters, we use
$\nu_{p,q}$ with $p+q \le 5$,
while we regard  $\nu_{p,q}$=0 for $p+q >6$.
This is legitimate because the charge fluctuation 
involving more than 5 carriers at a single site is quite small
for $n=2$ even in the uncorrelated case.
It is easily imagined that the Hund's-rule coupling would be
more effective than the case of nickel
because of the large number density of the carrier $n \ge 1$.

On the other hand, the density of states at the Fermi level
takes the relatively high value, as in nickel.
Another fact which is apparent in the band calculation
for cobalt and iron  is 
the double peak structure 
of the density of states.~\cite{rf:doublepeak}
The second peak which is away from the Fermi level 
can be effective in the presence of the electron correlation,
while, in the one-body approximation,
the condition for the ferromagnetism can be written down
only with the value at the Fermi surface.
To take these into account, we 
use the model density of states as in the previous section:
\begin{equation}
\label{eq:dos}
\rho(n)= \left\{
\begin{array}{ll}
\displaystyle{C_1}&
\qquad 1 \ge n \ge 4n_0 \\
\noalign{\vskip0.2cm}
\displaystyle{\frac{C_1}{s+(1-s)(\frac{n-n_0}{n_0})^2} }&
\qquad 4n_0 \ge n \ge 2n_0 \\
\noalign{\vskip0.2cm}
\displaystyle{\frac{C_1}{r+(1-r)(\frac{n-n_0}{n_0})^2} }&
\qquad 2n_0 \ge n \ge 0
\end{array} \right.  ,
\end{equation}
where $n_0$=0.2.
(See Fig.8 and Fig.9.)
The second peak at $n=3n_0$ 
(for $s^{-1} > 1$)
affects 
 the origin of the energy, $C_2$. (eq.~(\ref{eq:origin}).)
Consequently the presence of the second peak 
works against the ferromagnetism,
or spreads the incomplete ferromagnetic phase space 
into the complete ferromagnetism.
The phase diagrams are given in Figs.10,
11 and 12.
We see the effects of the Hund's-rule coupling and 
that of the second peak ($s^{-1} >1$).
Note that the Hund's-rule coupling ($j\sim 0.17$)
should not be large
compared with the case for nickel.
This is partly expected  by the fact that
the parameter $j$ is defined by $j\equiv J/W$,
and the band width $W$ ($\sim 5$eV)
 is larger for cobalt and iron 
than for nickel ($W\sim 4$eV).
(The wavefunctions for $d$-electrons  spread more than
those for Ni,
thus,
the values of $J$ and $I$ for
Fe and Co are more reduced than those for Ni.)
We also see that the phase boundary is sensitive 
to the Hund's-rule coupling $j$,
as expected
by the large carrier density.
The incomplete ferromagnetic region is enlarged 
by 
the effect of the second peak
as well as by the Hund's-rule coupling.
As for the reduction of the band width,
the factor $q \GA 0.9$ 
is given
in the region of our interest.
This is in accordance with the result for nickel:
the electron correlation is not strong.

To conclude, for the incomplete 
ferromagnetism to be realized in the realistic range of
$r^{-1}\sim 2$,
it is necessary
that
the Hund's-rule coupling should be $j\sim 0.17$
(see Fig.13)
and
the shape of the density of states need not 
have the same feature,
favorable for ferromagnetism, as for nickel.
Especially the double peak structure
can make the incomplete ferromagnetic state stable.
With respect to this point,
our conclusion notes that the 
ferromagnetism in cobalt and iron 
should not be taken to be
due to the high density of states at the Fermi level,
rather it is ascribed to the effect
of the Hund's-rule coupling.
With the help of the itinerant carrier motion,
the ferromagnetic correlation can be wide-spread and result in
the long-range ordering.
In these cases with more than one carriers per site,
the mechanism due to the polarity fluctuation 
assisted by the Hund's-rule coupling gives the clear
and natural explanation of the itinerant ferromagnetism.

\section{Summary and Discussions}
We generalized the Gutzwiller approximation 
to treat the realistic model of the itinerant ferromagnetic systems,
including the band degeneracy and the Hund's-rule coupling.
In deriving the required expression for
the reduction factor $q$ of the band width,
we found that
it can be interpreted intuitively that
$q$ is proportional to the square of
the sum of the multiple of the probability amplitudes
(square root of the probabilities). (eq.~(\ref{eq:ql}).)

We see that the Hund's-rule coupling $J$ 
is necessary 
to explain the 
itinerant ferromagnetism in the 3-$d$ transition metal.
The phase diagrams for $n=2$ showed
that the boundary between the paramagnetic and ferromagnetic states
is sensitive to the strength of the Hund's-rule coupling.
Specifically, $j=J/W$ should be around 0.17 for $n=2$.
In general, it is difficult to realize
the incomplete ferromagnetic state as the ground state,
because it needs to destabilize both of the
paramagnetic and the complete ferromagnetic states.
In effect, the energy difference between the paramagnetic
and ferromagnetic states is quite small,
so that any effect such as
the peculiarity of the band density of states
plays an important role
to determine the magnetism of the ground state.
As we saw above, the double peak structure
can affect the magnetism by the many-body effect, 
even though 
the density of states at the Fermi surface is 
fixed.

Concerning the reduction factor of the band width,
our result showed that the factor $q$ does not change 
appreciably from unity.
This is partly because of the
smallness of the interaction strength.
Moreover, we should note that our approximation is not 
accurate enough to be compared with the other method 
such as the diagram technique
and the t-matrix theory.~\cite{rf:Igarashi}
In fact, our method cannot
describe correctly the mass enhancement effect around the
Fermi surface,
for it is the dynamical effect due to 
the many-body correlation.
Therefore, it would not be proper here to 
argue on the discrepancy with respect to
the reduction of the band width
between the experiments and the result of the band 
calculation.~\cite{rf:band}
As it is the difficult task to treat the 
electron correlation in the band with degeneracy,
it is the future problem to 
improve upon the
method itself
to give more quantitative discussion.

\section*{Acknowledgments}

The author acknowledge Prof. K. Yamada for 
fruitful discussions.
This work is supported  by
the Grant-in-Aid for 
Scientific Research from the Ministry of Education,
Science and Culture.


\vskip 2cm

\noindent {\bf Figure}

\begin{itemize} 
\item{ Fig. 1 \ : \ \ \ 
Examples of the hopping process.
}
\item{ Fig. 2 \ : \ \ \
$\rho$ as a function of $n$. eq. (4.1))
}
\item{ Fig. 3 \ : \ \ \
$\rho$ as a function of $\epsilon$.}
\item{ Fig. 4 \ : \ \ \
The phase diagram for $j$=0.3.}
\item{ Fig. 5 \ : \ \ \
The phase diagram for $j$=0.15.}
\item{ Fig. 6 \ : \ \ \
The phase diagram for $j$=0.}
\item{ Fig. 7 \ : \ \ \
Magnetic moment as a function of $n$.}
\item{ Fig. 8 \ : \ \ \
$\rho$ as a function of $n$. (eq. (5.1))}
\item{ Fig. 9 \ : \ \ \
$\rho$ as a function of $\epsilon$.}
\item{ Fig. 10 \ : \ \ \
The phase diagram for $s^{-1}$=1 and $j=0.17$.}
\item{ Fig. 11 \ : \ \ \
The phase diagram for $s^{-1}=r^{-1}/1.2$ and $j=0.17$.}
\item{ Fig. 12 \ : \ \ \
The phase diagram for $s^{-1}=r^{-1}/1.2$ and $j=0.18$.}
\item{ Fig. 13 \ : \ \ \
The phase diagram for $r^{-1}=2$,
$s^{-1}=r^{-1}/1.2$}
\end{itemize}


\begin{thebibliography}{99}
\bibitem{rf:cite} In effect, we would see that the 
calculation gives
different factors, $q_{-}$ and $q_{+}$, for each of $<\Psi|
aa|\Psi>$
and $<\Psi|
a^{\dagger}a^{\dagger}|\Psi>$.
However, we can introduce only one parameter $q=\sqrt{q_{-}q_{+}}$
for the physically relevant models in which
the annihilation and the creation processes 
appear pairwise.
\bibitem{rf:cite2} In figures, we showed the phase boundary between
the incomplete and complete ferromagnetic states.
The boundary, however, does not macroscopically represent the 
phase transition line because there is no
broken symmetry between them.
\end{thebibliography}
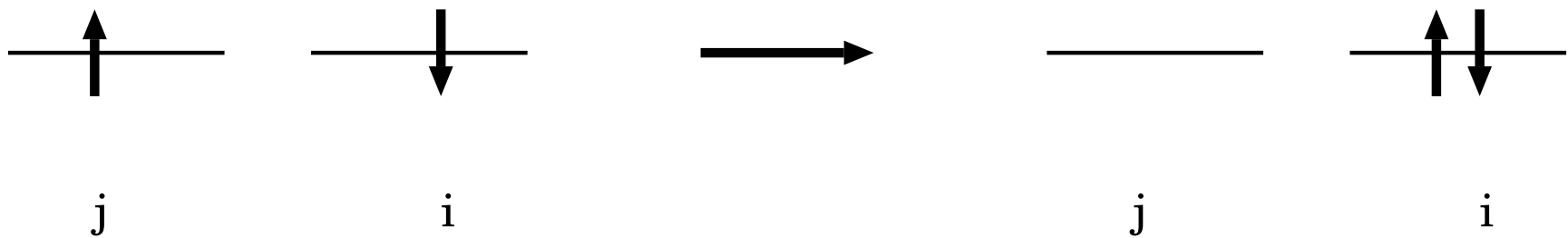
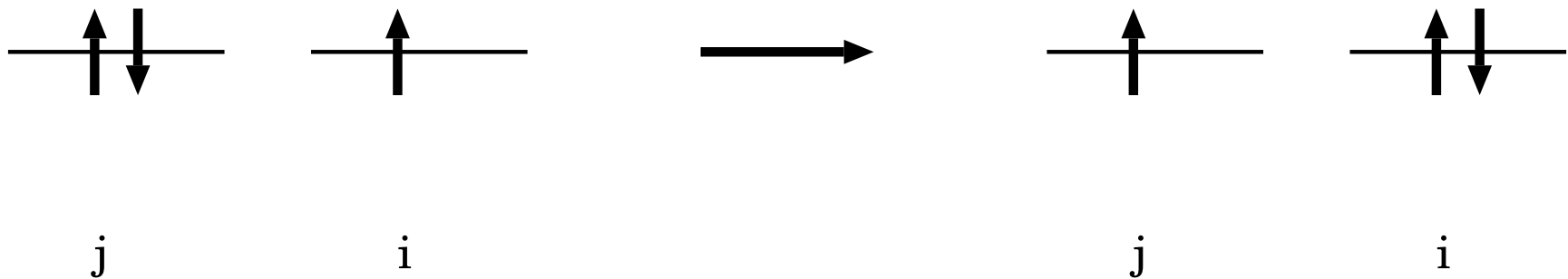
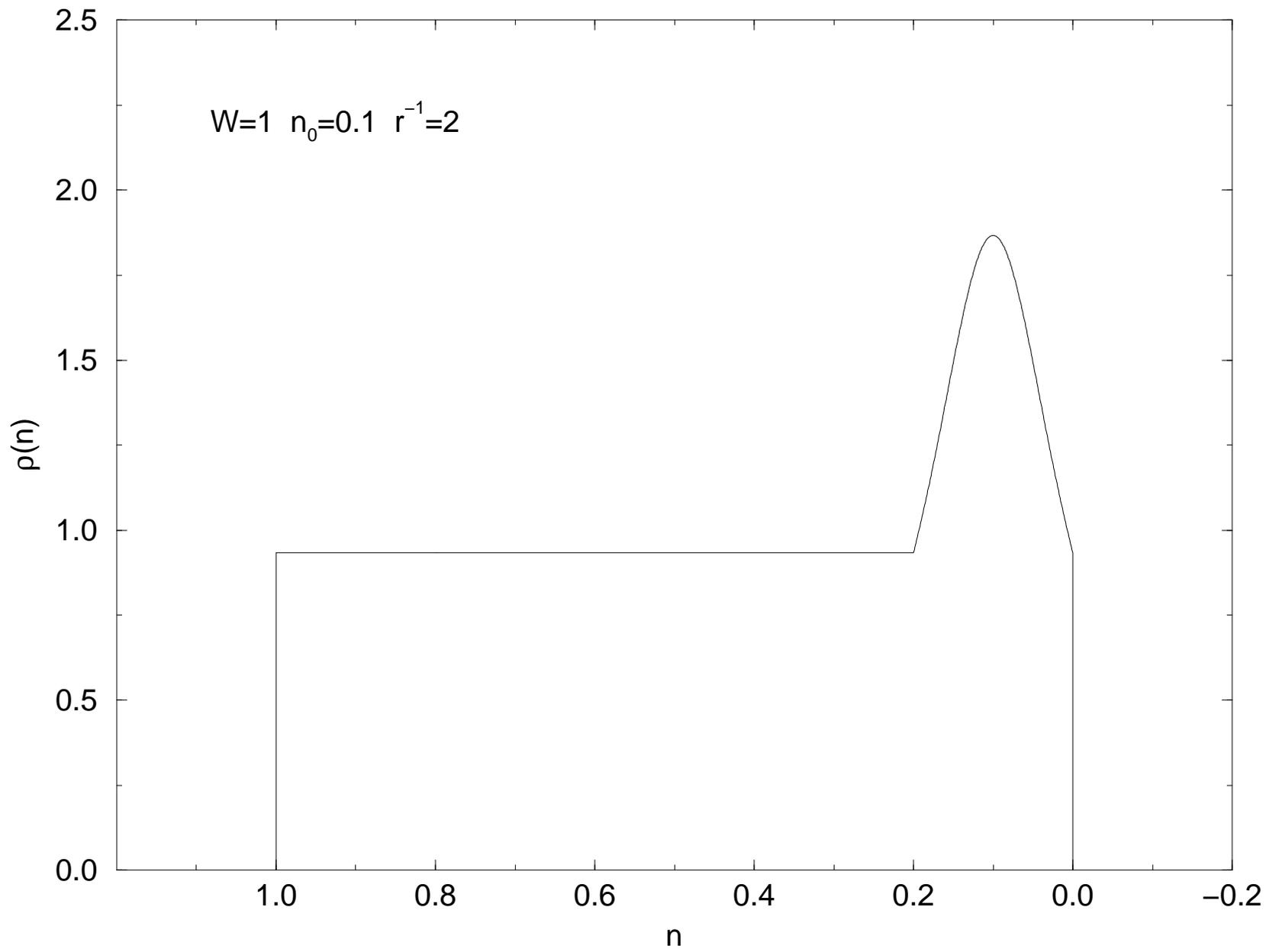
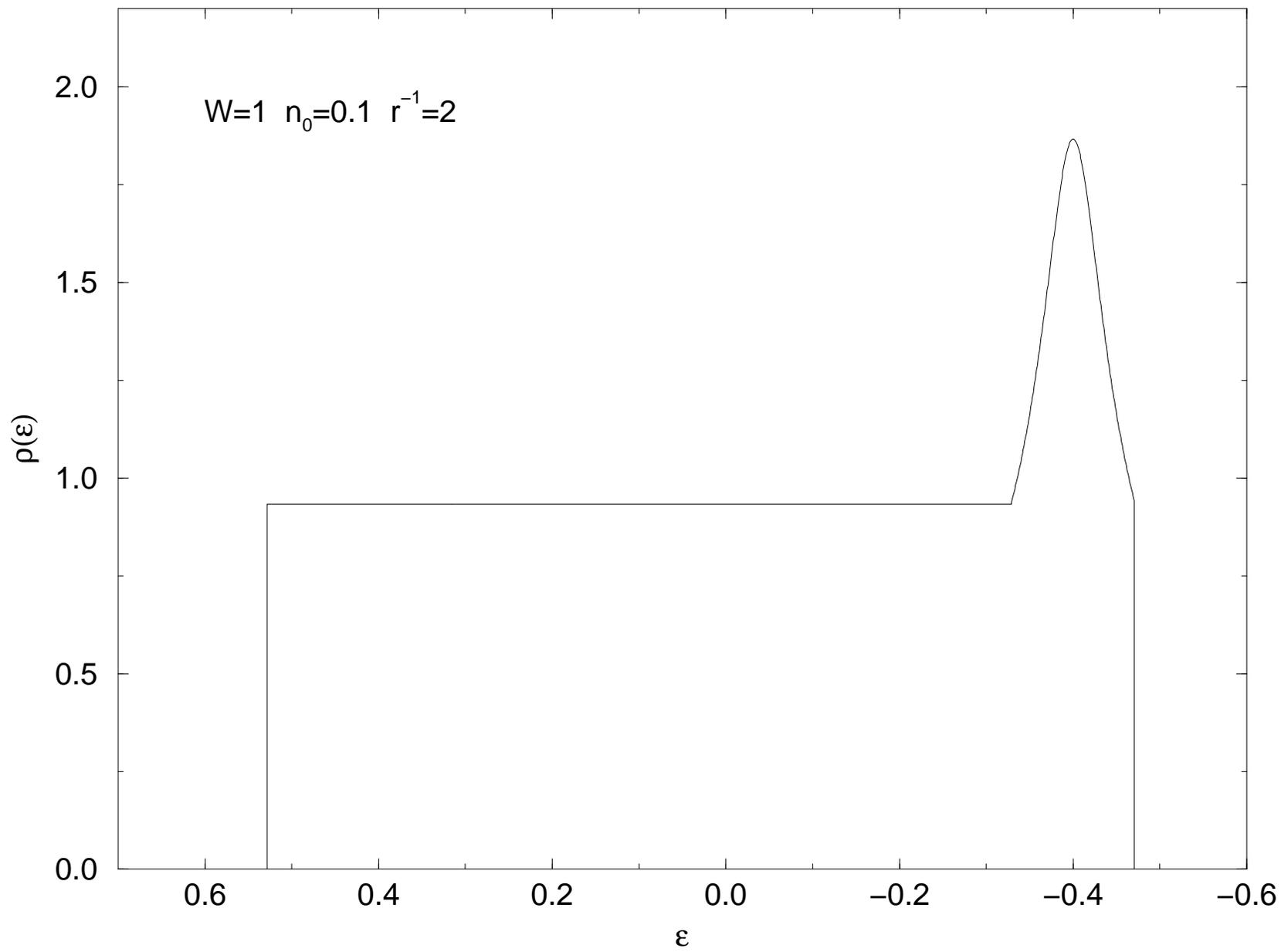
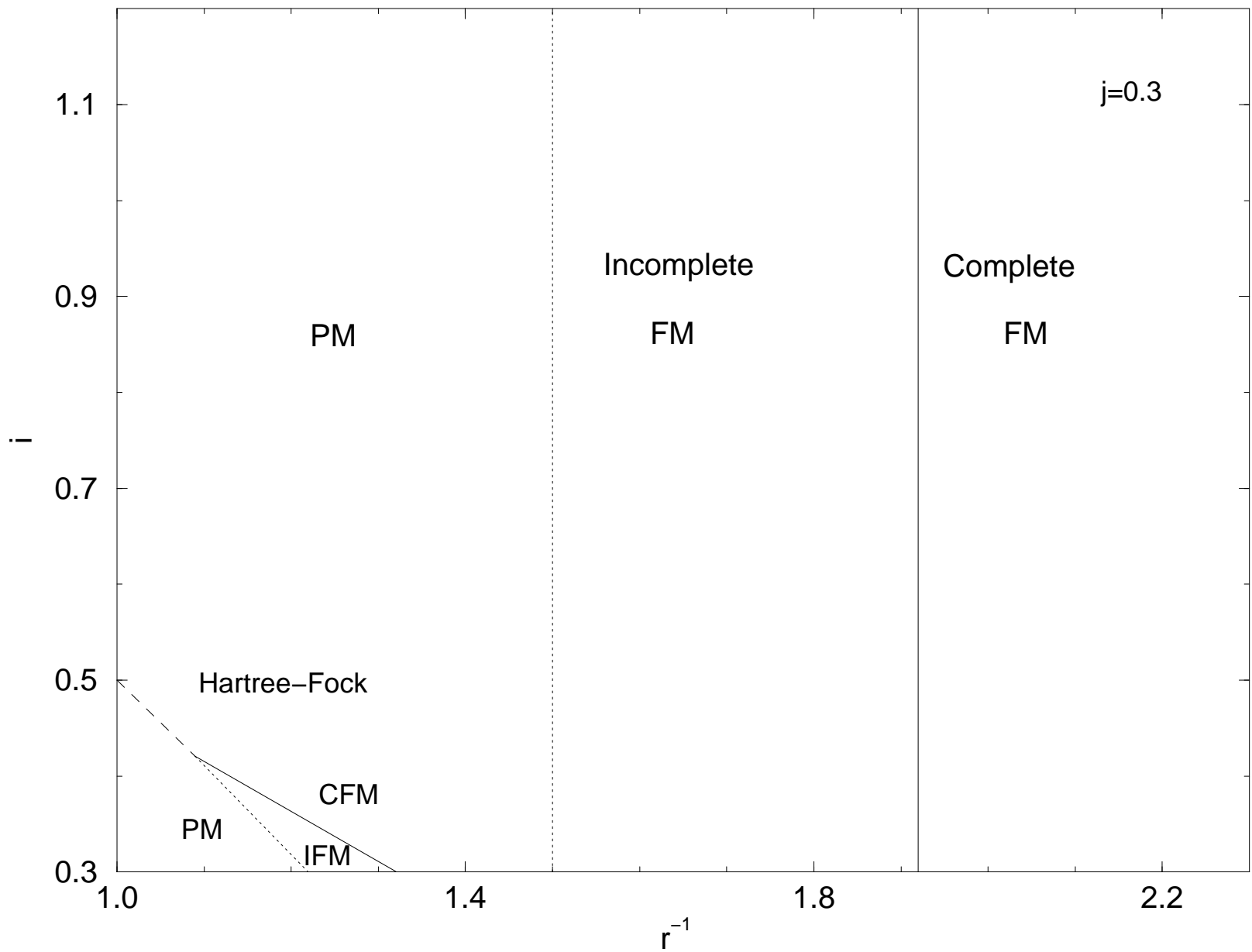
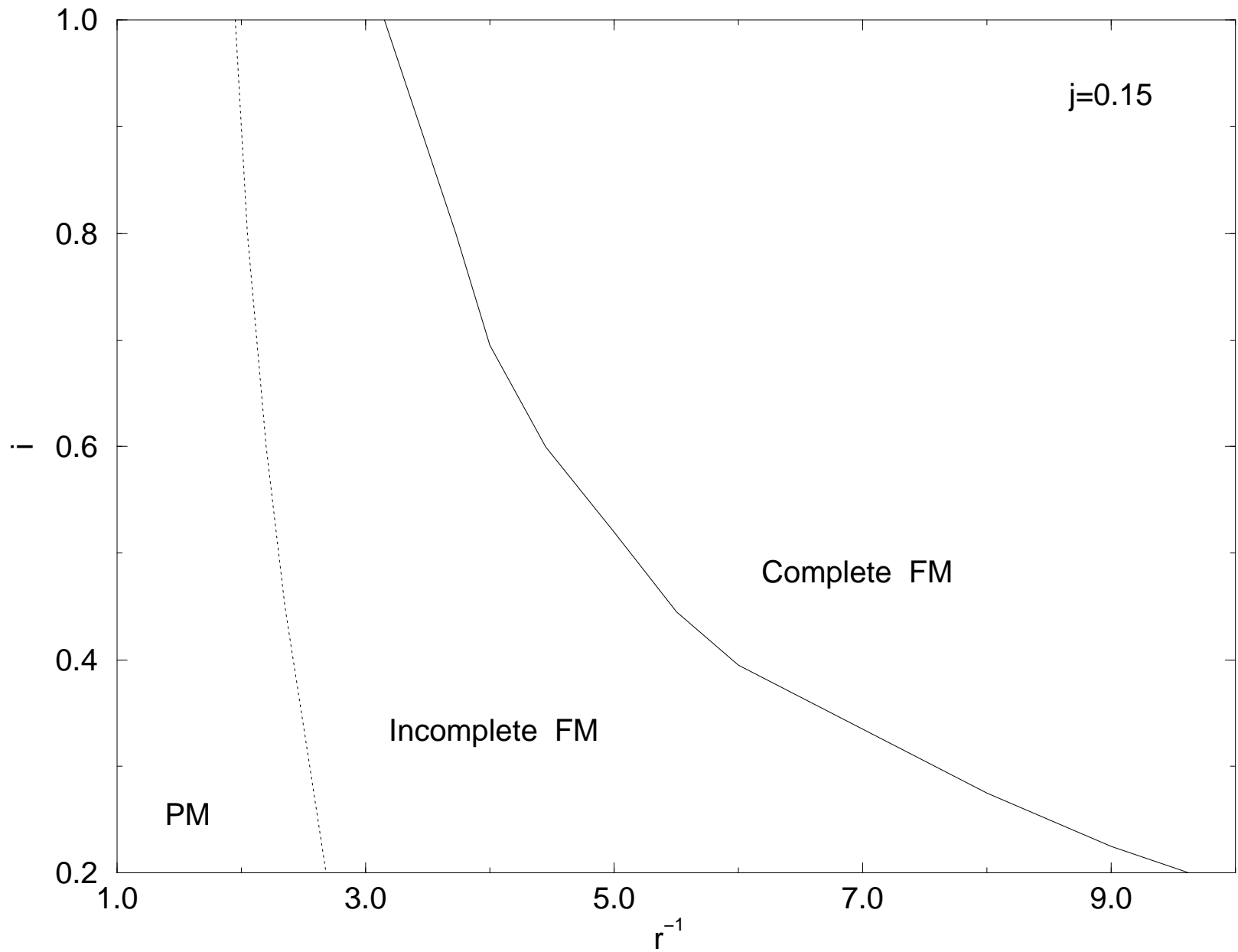
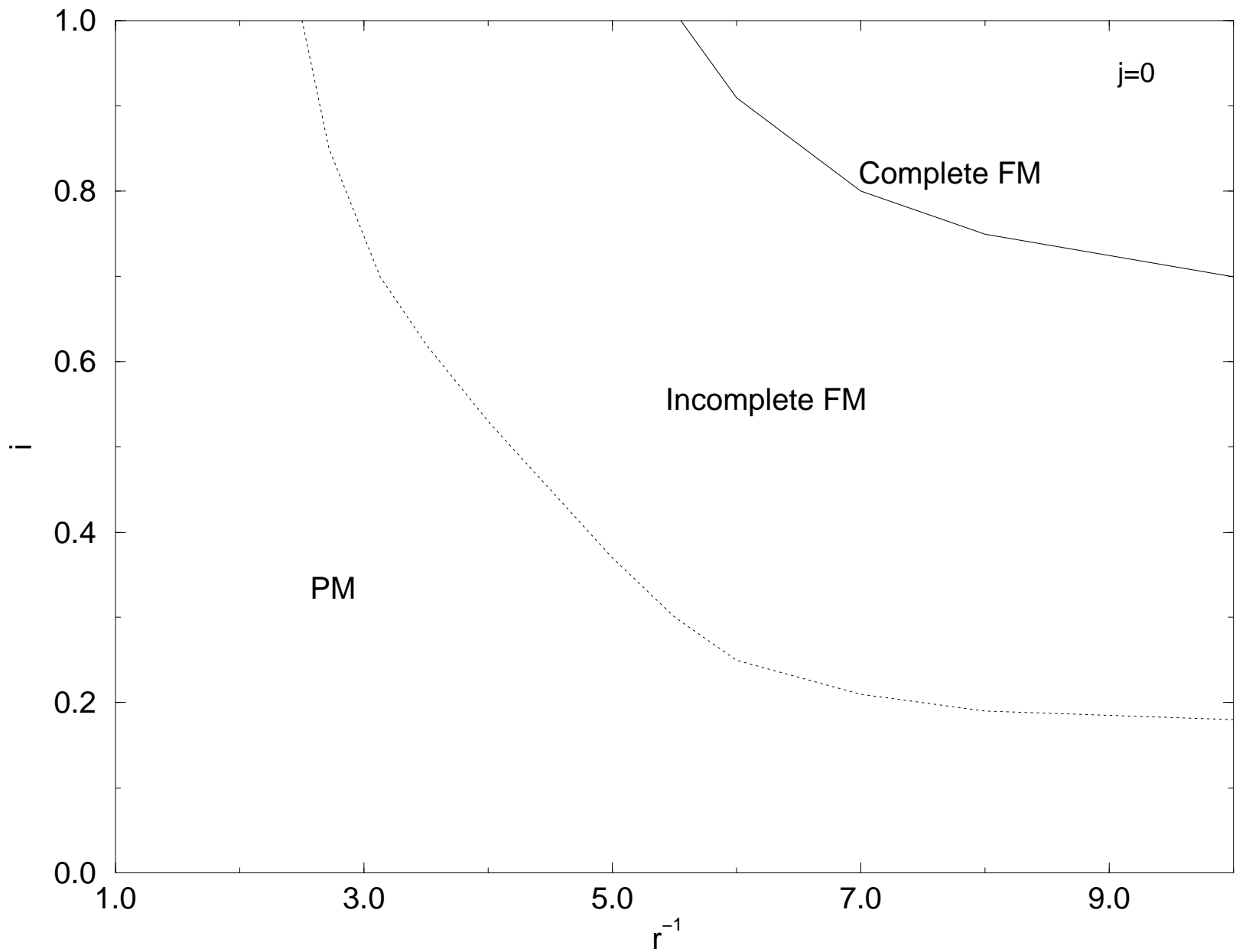
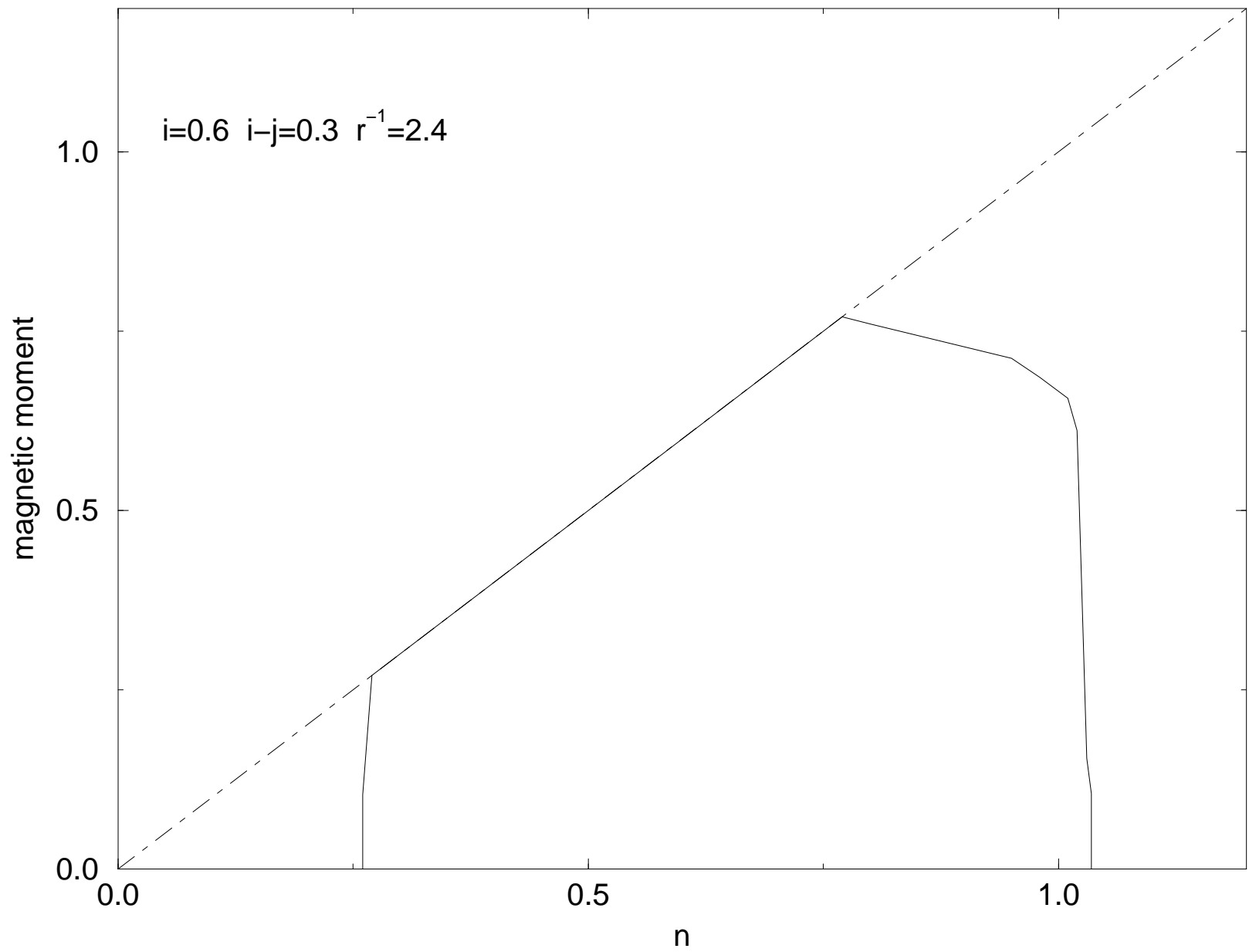
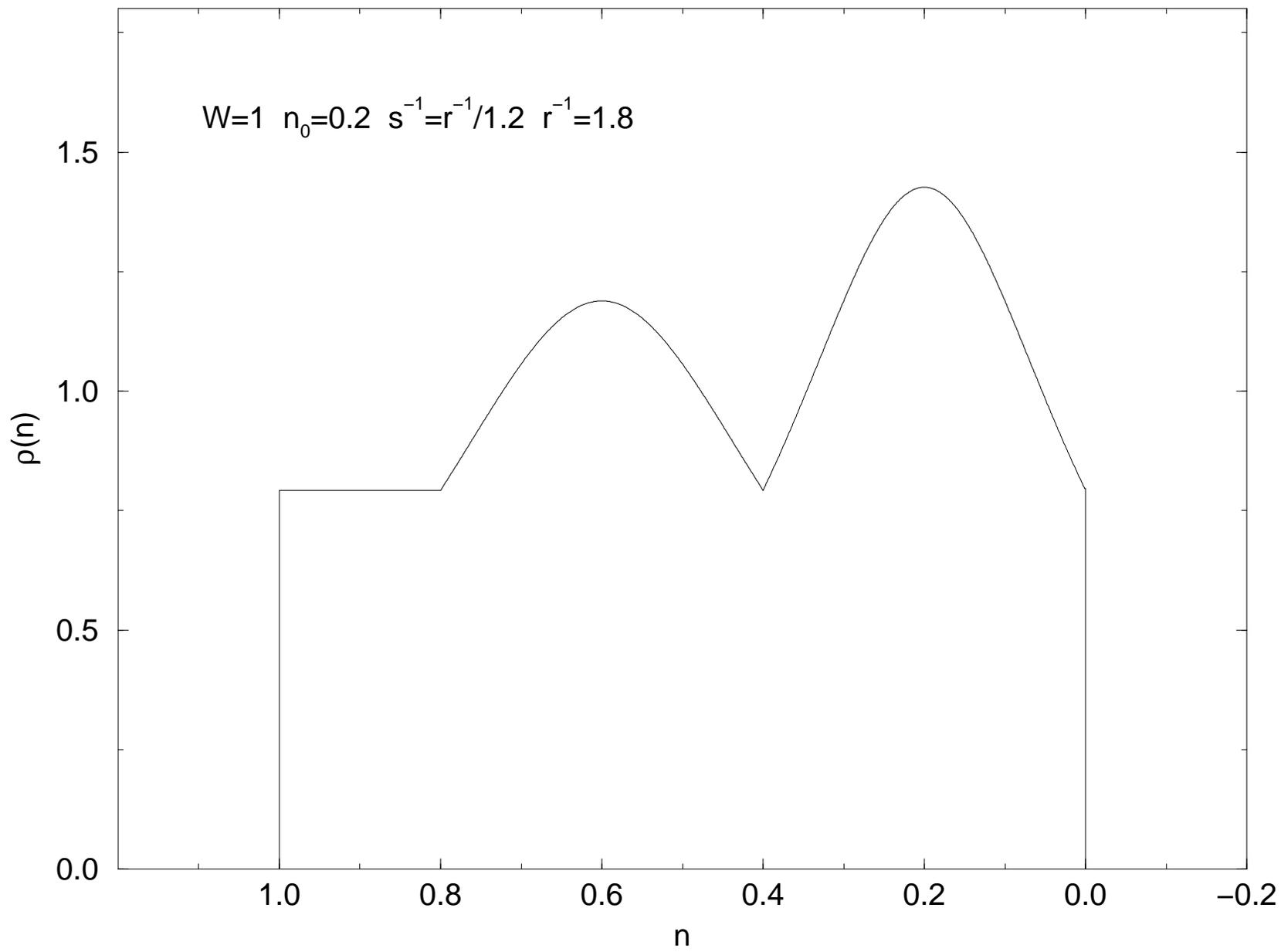
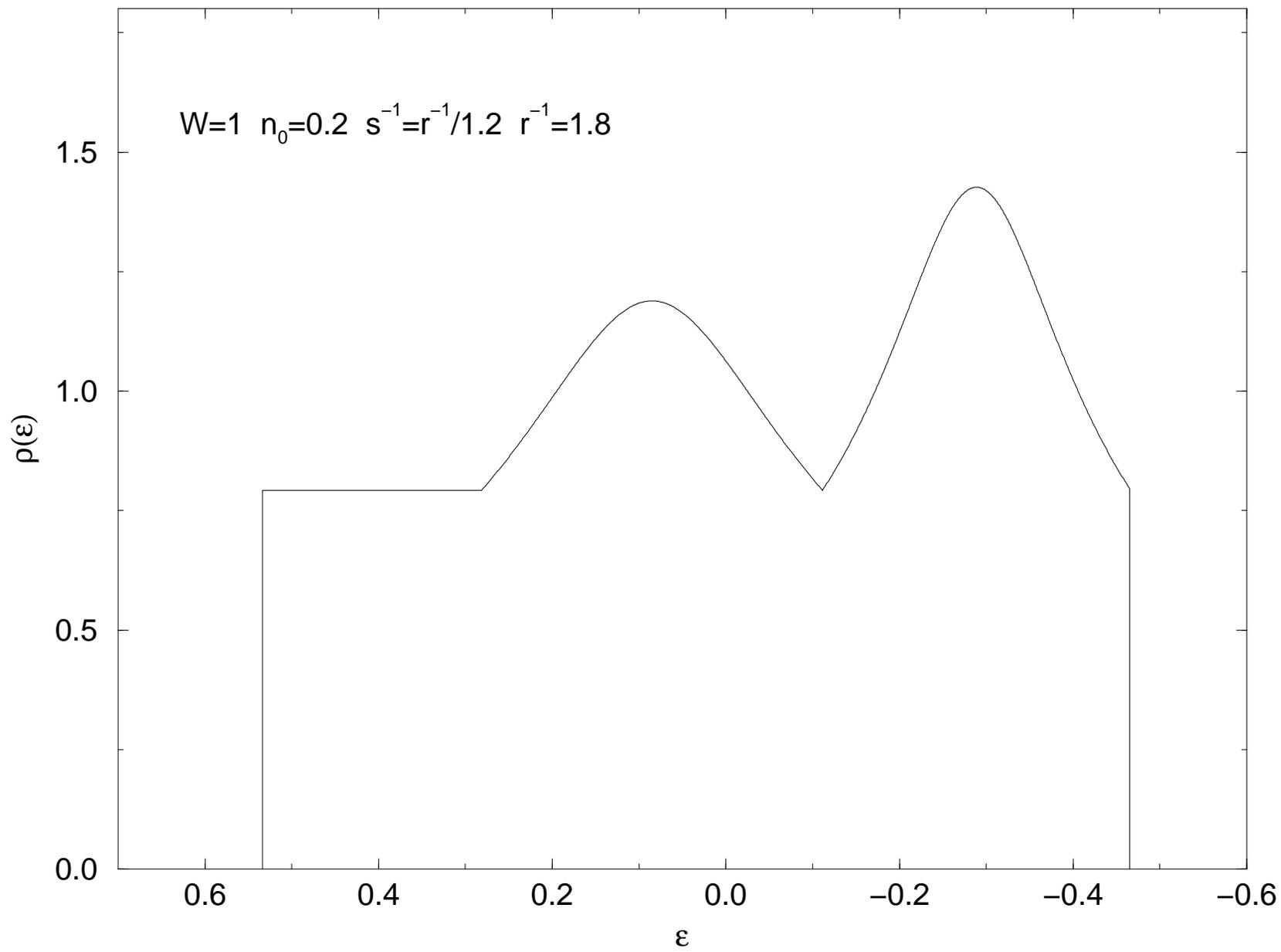
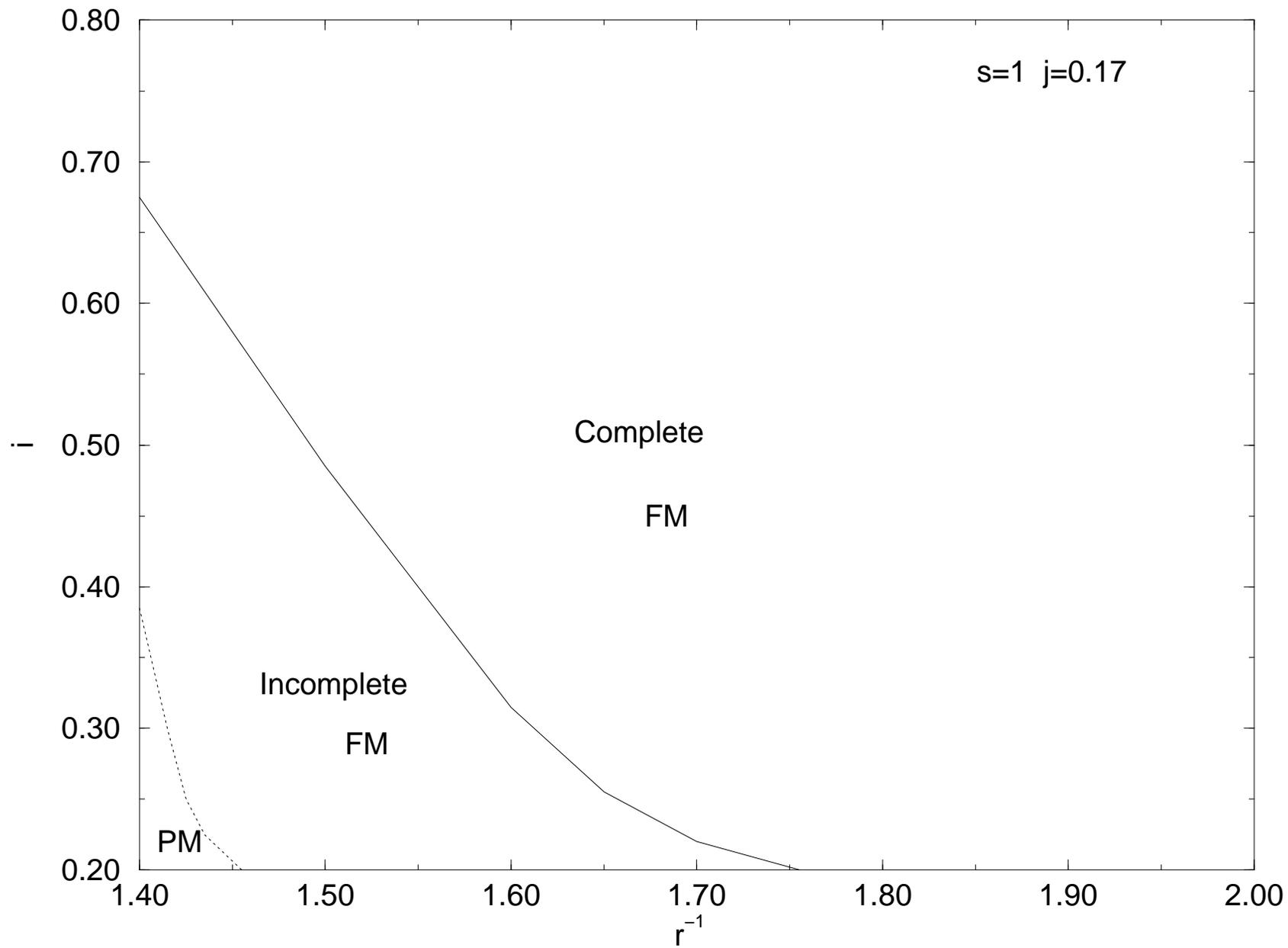
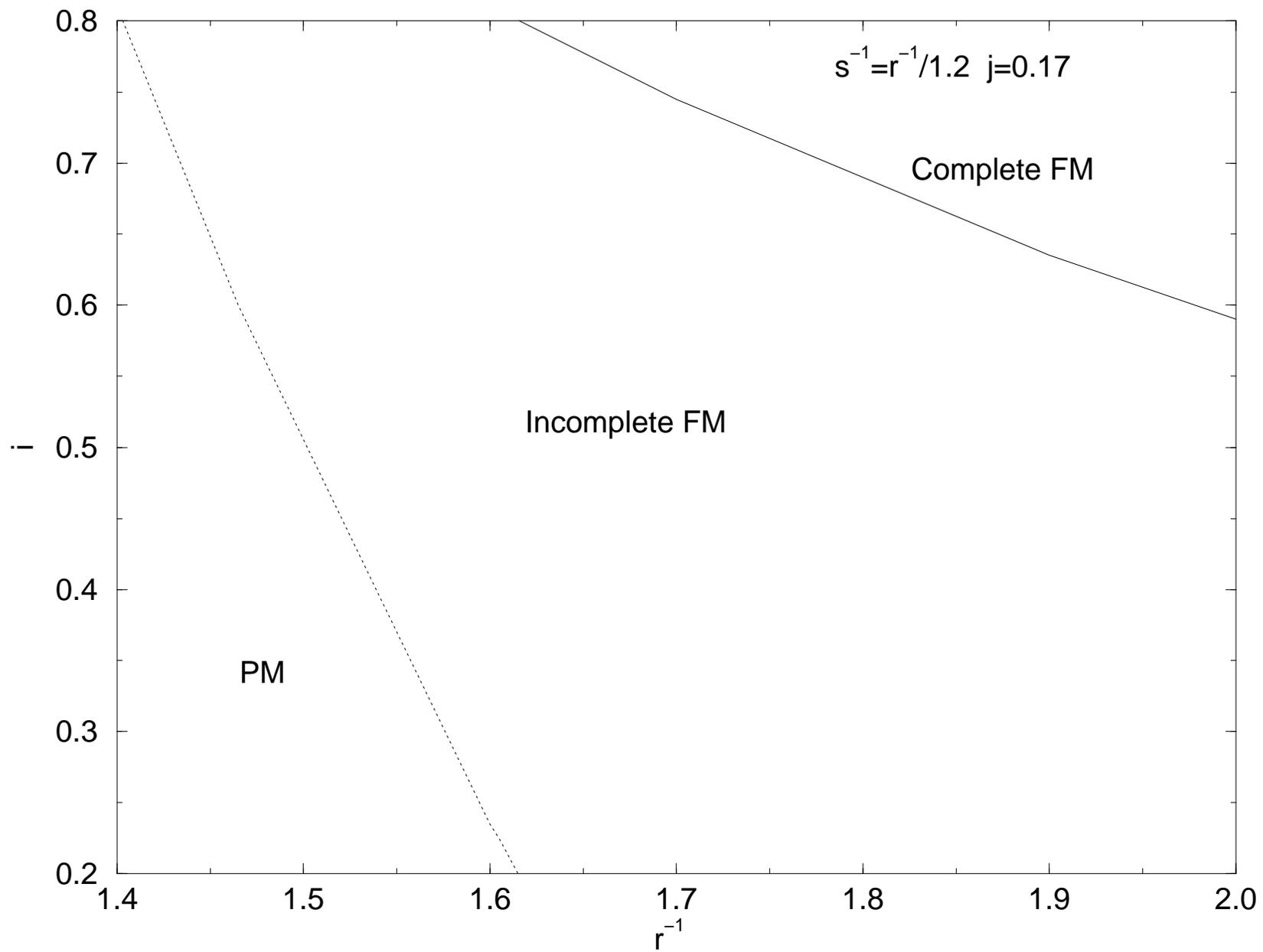
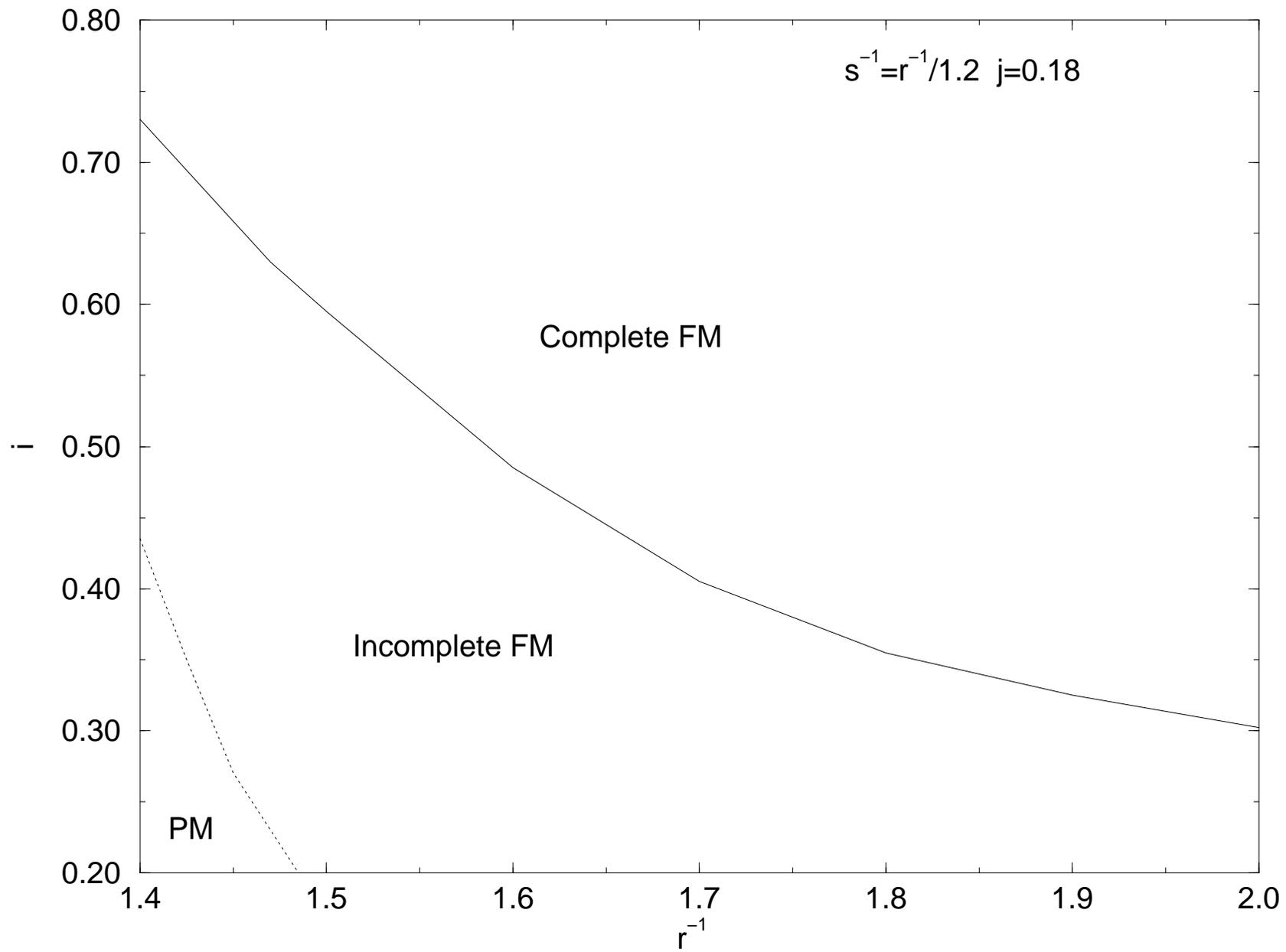
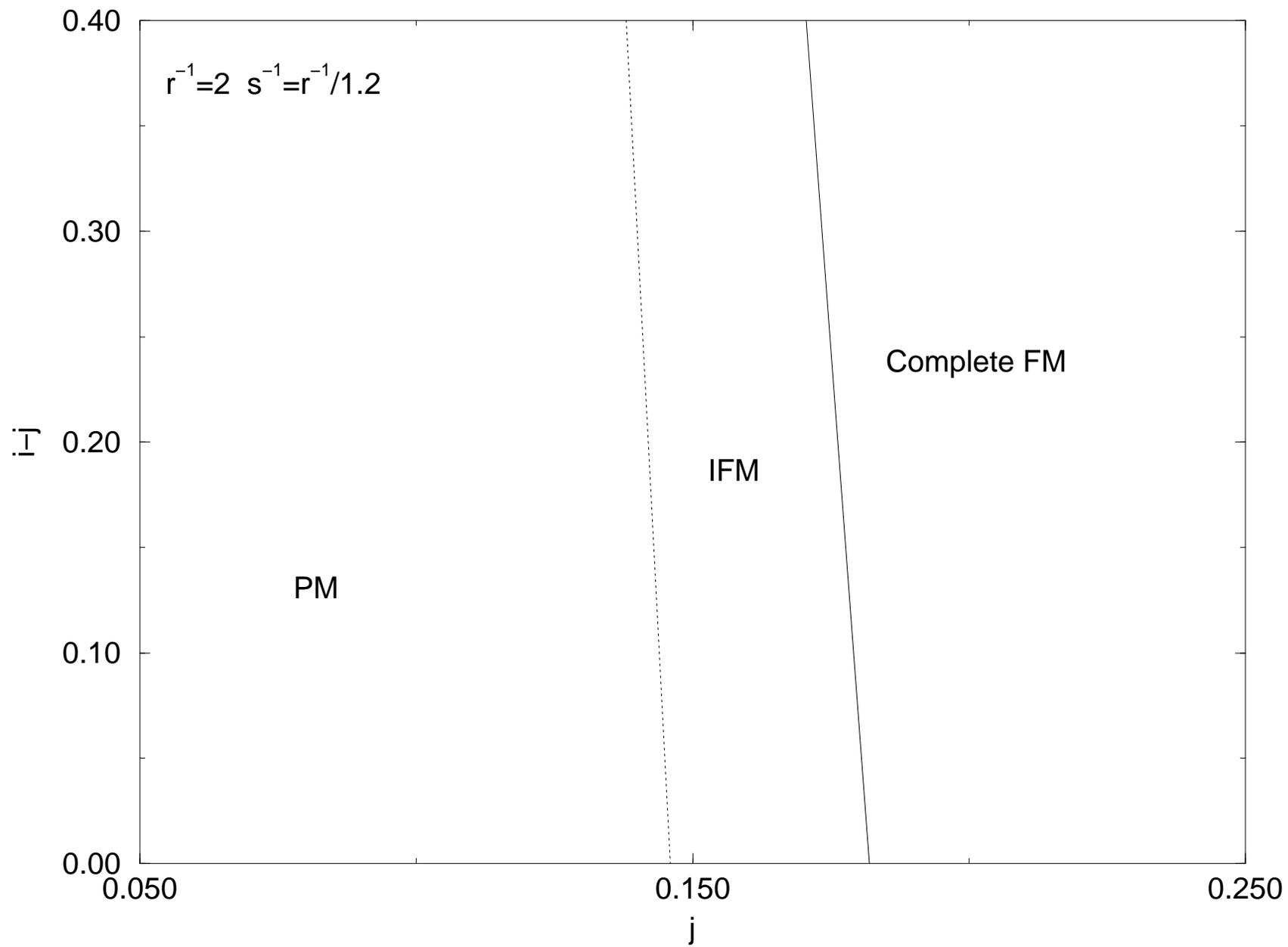
\end{document}